\documentclass[10pt,b5paper,twoside]{padeu}  
\usepackage{epsfig,natbib_padeu}             

\begin{document}

\title{Celestial positions in radio and optical}
\author{S. Frey\inst{1},
	P. Veres\inst{2},
	K. Vida\inst{2}}
\authorrunning{S. Frey, P. Veres, K. Vida}
\institute{\inst{1} F\"OMI Satellite Geodetic Observatory, P.O. Box 585, H-1592 Budapest, Hungary \\
	   \inst{2} E\"otv\"os University, Department~of Astronomy, P.O. Box 32, H-1518 Budapest, Hungary \\
}
\email{frey@sgo.fomi.hu}


\abstract{We discuss the importance of the direct link between the most accurate radio and optical reference frames that will become possible with the next-generation space astrometry missions in about a decade. The positions of more than 500 active galactic nuclei that are common in the precise Very Long Baseline Interferometry (VLBI) catalogues and the Sloan Digital Sky Survey (SDSS) Data Release 4 (DR4) are compared. While obtaining an ``independent'' estimate for the SDSS coordinate accuracies, we find indications that the assumption of spatially coincident brightness peaks for the same objects in radio and optical does not hold for each object.
}

\maketitle


\section{Introduction}

From 1998, the International Astronomical Union (IAU) adopted the International Celestial Reference Frame (ICRF)
as the fundamental celestial reference frame \citep{ma98}. The ICRF is defined by the positions of 212 compact extragalactic {\it radio} sources (active galactic nuclei, AGNs) regularly observed with Very Long Baseline Interferometry
(VLBI) over a long period of time. In the {\it optical}, the HIPPARCOS catalogue \citep{perr97} is the most accurate available to date. It is linked to the ICRF with a variety of observing techniques but mainly through relative astrometry with respect to nearby radio-loud AGNs using radio interferometric observations of twelve radio stars \citep{kova97}. The coordinate axes were aligned with the extragalactic radio frame to within $\pm0.6$ milli-arcseconds (mas) at the epoch 1991.25, with a non-rotation within $\pm0.25$~mas/yr. The quality of this link is known to degrade with time, due to e.g. the uncertainties in the measured stellar proper motions.

With a sensitive next-generation space astrometry mission -- like the European Space Agency's Gaia spacecraft to be lauched in 2011 \citep{perr05}, or the U.S. Space Interferometry Mission (SIM) PlanetQuest \citep{marr03} -- a quasi-inertial reference frame can directly be established in the optical as well. Gaia will detect $\sim$500~000 quasars brighter than the limiting magnitude $G=20^{\rm{m}}$. Its celestial reference frame will be defined by a sample of at least $\sim10$~000 clearly detected quasars expected to be observed during the mission \citep[e.g.][]{mign02}. In terms of precision, the Gaia optical reference frame will supersede the current radio ICRF. However, the latter will retain its importance at least because the Earth rotation and orientation are uniquely measured with VLBI by linking the terrestrial reference frame defined by the radio antenna locations and the quasi-inertial
celestial reference frame defined by the extragalactic radio sources \citep{fey04}.


\section{Direct link between reference frames}

The space- and ground-based optical astometric sky surveys conducted to date are not sensitive enough to detect the faint optical counterparts of the distant radio-loud AGNs routinely observed with VLBI. An extensive astometric program is going on at the U.S. Naval Observatory to construct an optical extragalactic reference frame \citep[e.g.][]{zach03,assa03}.

The limiting visual magnitude of Gaia will enable observing the optical couterparts
of practically all the ICRF defining radio-loud AGNs \citep[their median visual magnitude is $V=18.1$,][]{fey01}, allowing in principle an accurate {\it direct} link between the radio and optical reference frames. This link is essential not only for astrometry,
but for astrophysical applications as well. The high-resolution stucture of the AGNs observed at
different electromagnetic wavelengths can only be registered correctly if the coordinates
are expressed in consistent systems. The quality of such a link is enhanced with the increasing number of common
objects in both the optical and the radio. Any survey that offers an extensive list
of suitable link sources is valuable for improving the tie between the radio and optical reference frames,
even if the individual positional accuracy of the sources is somewhat worse than that of
the reference frame defining objects. Once completed, e.g. the Deep Extragalactic VLBI--Optical Survey \citep[DEVOS,][]{moso06} could provide a sizeable set of additional useful link sources \citep{frey05}.


\section{A case study: SDSS coordinates}

\begin{figure}[h]
   \centering
   \includegraphics[width=9cm,bb= 80 365 550 704,clip= ]{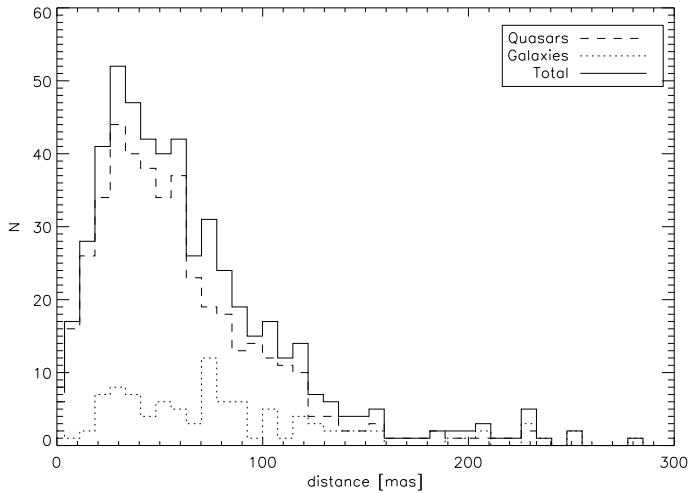}
   \caption{Histogram of optical--VLBI radial position differences for 524 ICRF and/or VCS sources that are also identified in SDSS DR4 as quasars or galaxies. The distribution is consistent with $\sim60$~mas ($1\sigma$) accuracy in both right ascension and declination in SDSS.}
   \label{frey:fig1}
\end{figure}

The imaging catalogue of the $4^{\rm th}$ Data Release\footnote{\tt http://www.sdss.org/dr4}
\citep[DR4,][]{adel06} of the Sloan Digital Sky Survey (SDSS) covers almost 6700 square degrees of the sky, mainly on the Northern hemisphere and around the Equator. Although SDSS is of course not an astrometric sky survey, the positions of its objects are thoroughly calibrated using the USNO CCD Astrograph Catalogue \citep[UCAC,][]{zach00} and the Tycho-2 catalogue \citep{hog00}, with a declared absolute rms accuracy better than 100~mas per coordinate \citep{pier03}. The large sky coverage and the $R\sim22^{\rm{m}}$ magnitude limit of SDSS make possible to identify the counterparts of many radio-loud AGNs that have accurate radio positions available.
We compared the ICRF \citep{ma98,fey04} and the VLBA Calibrators Survey\footnote{\tt http://www.vlba.nrao.edu/astro/calib}
\citep[VCS,][]{beas02,foma03,petr05,petr06} catalogues with the SDSS DR4. We looked for optical cross-identifications of radio sources that have 10~mas formal positional accuracy or better in both right ascension ($\alpha$) and declination ($\delta$) coordinates. (The average values are actually below 1~mas.) This way the source positions in the sample are at least an order of magnitude more accurate than in the optical. Among the all-sky set of 2628 ICRF and VCS sources satifying the above criteria, optical counterparts of 524 were found in the SDSS within a search radius of 300~mas.
The majority of our sources are in the range $7^{\rm h}<\alpha<18^{\rm h}$ and $-5^{\circ}<\delta<+70^{\circ}$, with a few additional sources well distributed in right ascension at $-15^{\circ}<\delta<+15^{\circ}$.
The distribution of the SDSS--VLBI position differences for these objects is shown in a histogram in Fig.~\ref{frey:fig1}.

One may suspect that the ``outliers'' seen at the tail of the distribution that represent SDSS--VLBI cross-identifications well beyond the expected radial distance are {\it chance coincidences}. To test this hypothesis, we constructed four ``false'' radio source lists by shifting the right ascension or declination coordinates of all the ICRF sources by a large amount ($\pm1^{\circ}$). We then tried to find SDSS optical counterparts within 300~mas radius for these ``objects'' (over 10~000 in total). Only two chance coincidences were found, suggesting that most if not all of our outliers (21 AGNs with optical--radio separation larger than 180~mas) in Fig.~\ref{frey:fig1} are {\it real identifications}. This result can be considered as a sign of warning. When comparing directly the radio and optical positions, we naturally assume that the optical and radio emission peaks physically coincide. Because the activity of these AGNs is driven by their central supermassive black holes, this assumption seems plausible in general. However, there are objects for which it is not necessarily true. Such objects should be avoided when the radio and optical reference frames are linked. (On the other hand, these AGNs may well be interesting from an astrophysical point of view.)
Although the optical images -- at least for the quasars -- are unresolved, earlier studies of ICRF radio sources with extended ($\sim10$ mas) VLBI structure also found evidence of non-coincidence between the radio and optical centers \citep{dasi02}.

The distribution in Fig.~\ref{frey:fig1} is generally consistent with $\sim60$~mas positional uncertainty per coordinate, in a good agreement with the values determined for SDSS \citep{pier03}. The {\it optical minus radio} right ascension ($\Delta\alpha \cos\delta$) differences have $\sigma_{\Delta\alpha}=58.3$~mas standard deviation and a negligible $\langle \Delta\alpha \cos\delta \rangle=0.4$~mas mean value. While the declination differences have similar standard deviation ($\sigma_{\Delta\delta}=59.9$~mas), there appears to be an average offset of $\langle \Delta\delta \rangle=12.3$~mas. As expected, the coordinate differences for optically resolved SDSS galaxies (102 in the sample) show larger standard deviation ($\sigma\approx80$~mas) than for the quasars alone ($\sigma\approx53$~mas).

To see if there is a set of rotations which could transform the system defined by the SDSS optical positions to the radio system, a three-parameter least-squares adjustment was performed. (The uncertainties of the individual points were not considered here.) The estimates for the rotation angles around two of the Cartesian coordinate axes ($x$: $\alpha=0$, $\delta=0$ and $z$: $\alpha=0$, $\delta=\pi/2$) were insignificant: ($-5.7\pm3.1$)~mas and ($3.0\pm2.7$)~mas, respectively. The ($14.6\pm2.4$)~mas rotation obtained around the $y$ axis ($\alpha=\pi/2$, $\delta=0$), although small compared to the individual coordianate uncertainties, may be an indication for a small systematic difference in the SDSS  positions with respect to the ICRF. Notably, a $-15$~mas mean declination offset (i.e. similar in extent but with an opposite sign) was found by \citet{assa03}, who determined the UCAC optical positons of 172 ICRF sources in the range $-30^{\circ}<\delta<+25^{\circ}$ and compared them with the radio positions. In our case, after transforming the SDSS coordinates using the rotation angles estimated above, the mean optical--radio differences became small: $\langle \Delta\alpha \cos\delta \rangle=-0.3$~mas and $\langle \Delta\delta \rangle=3.2$~mas.


\section{Conclusions}
SDSS, although not an astrometric survey, could be used to compare optical and radio coordinates of 524 AGNs, all with  accurate positions determined with VLBI at the mas or sub-mas level. We confirmed that the SDSS coordinates of these objects agree in both right ascension and declination with $\sigma\approx60$~mas. There may be a small ($\sim15$~mas) systematic rotation between the reference frame of the SDSS and the ICRF. We found that the radio and optical brightness peaks do not necessarily coincide in the sky for each object. It is therefore essential to identify these ``outliers'' before high-precision refererence frame connections are made in the future.


\begin{acknowledgement}
We thank S. Nagy and G. Vir\'ag for helpful discussions. This work was supported by the Hungarian Scientific Research Fund (OTKA T046087). S.F. acknowledges the Bolyai Research Scholarship received from the Hungarian Academy of Sciences.
Funding for the SDSS has been provided by the Alfred P. Sloan Foundation, the Participating Institutions, the National Science Foundation, the U.S. Department of Energy, the National Aeronautics and Space Administration (NASA), the Japanese Monbukagakusho, the Max Planck Society, and the Higher Education Funding Council for England.
The VLBA Calibrators list was generated from observations and reductions by the NASA Goddard Space Flight Center VLBI group, the U.S. National Radio Astronomy Observatory and the U.S. Naval Observatory.
\end{acknowledgement}


\end{document}